\documentclass[11pt]{article}

\usepackage{acl}

\usepackage{times}
\usepackage{latexsym}

\usepackage[T1]{fontenc}

\usepackage[utf8]{inputenc}

\usepackage{microtype}

\usepackage{inconsolata}

\usepackage{graphicx}

\usepackage{tabularx}
\usepackage{subcaption}
\usepackage{xcolor}
\usepackage{booktabs}
\usepackage{array}

\newcolumntype{Y}{>{\raggedleft\arraybackslash}X}

\usepackage{algorithm}
\usepackage{algorithmicx}
\usepackage{algpseudocode}
\usepackage{float}
\usepackage{multirow}
\usepackage{amsmath}
\usepackage{amssymb}

\usepackage{makecell,siunitx,booktabs}
\newcommand{\note}[1]{\textcolor{blue}{\textbf{TODO: #1}}}

\usepackage[most]{tcolorbox}
\usepackage{xcolor}
\usepackage{lipsum}

\definecolor{myorange}{HTML}{B45F04}
\definecolor{mycream}{HTML}{FFF8EE}

\tcbset{
  promptstyle/.style={
    enhanced,
    colback=mycream,
    colframe=myorange,
    fonttitle=\bfseries\footnotesize,     
    coltitle=white,
    title filled,
    rounded corners,
    boxrule=0.5pt,                        
    attach boxed title to top left={yshift=-0.8mm, xshift=1mm},
    boxed title style={
      colback=myorange,
      boxrule=0pt,
      rounded corners,
    },
    sharp corners=downhill,
    before skip=3pt,                      
    after skip=3pt,
    left=2pt, right=2pt, top=2.5pt, bottom=2.5pt,  
    colbacktitle=myorange!90!black,       
    fontupper=\small,                     
  }
}

\newcolumntype{R}[1]{>{\raggedleft\let\newline\\\arraybackslash\hspace{0pt}}m{#1}}

\title{Context-Guided Decompilation: A Step Towards Re-executability}

\author{
  Xiaohan Wang \\
  Vanderbilt University \\
  Nashville, TN, USA \\
  \texttt{} \\
  \And
  Yuxin Hu \\
  Vanderbilt University \\
  Nashville, TN, USA \\
  \texttt{} \\
  \And
  Kevin Leach \\
  Vanderbilt University \\
  Nashville, TN, USA \\
}  

\begin{document}
\maketitle
\begin{abstract}
Binary decompilation, which aims to recover source code from compiled binaries, can be viewed as a low-resource and high-constraint form of neural machine translation.
However, existing decompilation techniques often fail to produce source code that can be successfully recompiled and re-executed, particularly for binaries that have undergone compiler optimizations. Recent advances in large language models (LLMs) have enabled neural approaches to decompilation, but the generated code is typically only semantically plausible rather than truly executable, limiting their practical usability. 
These shortcomings arise from compiler optimizations and the loss of semantic cues in compiled code, which LLMs struggle to recover without contextual guidance.
To address this challenge, we propose ICL4Decomp, a hybrid decompilation framework that leverages in-context learning (ICL) to guide LLMs toward generating re-executable source code. 
Notably, our approach is model-agnostic and requires no fine-tuning, enabling plug-and-play use with off-the-shelf LLMs.
We evaluate ICL4Decomp across multiple datasets, optimization levels, and compilers, demonstrating around 30\% improvement in re-executability over state-of-the-art neural decompilation methods while maintaining robustness.
\end{abstract}

\section{Introduction}
Binary decompilation can be conceptualized as a specialized instance of neural machine translation (NMT): translating low-level binaries into high-level source code.
It plays a critical role in security auditing, vulnerability analysis, and malware reverse engineering, particularly when the original source code is unavailable~\cite{cifuentes2001computer,yakdan2016helping}. 
Despite decades of research, producing readable and executable source code from optimized binaries remains a challenging problem.

Unlike traditional natural language translation, 
the fundamental difficulty of binary translation lies in the fact that compilation is a lossy process~\cite{wilhelm2013compiler}. During compilation, high-level semantic information, such
as variable names, control structures, and type annotations, is discarded or transformed~\cite{cao2023revisiting,yang2025human}. 
This loss is exacerbated by aggressive compiler
optimizations, which further obscure data flow and control structure, making accurate
decompilation increasingly difficult.

Traditional decompilers rely on heuristic rules and control-flow analysis to reconstruct
source-level structures~\cite{IDAPro,2025NationalSecurityAgencyGhidra}. While effective for unoptimized binaries,
these tools often struggle under higher optimization levels, frequently producing output
that is incomplete, misleading, or not recompilable. 
As a result, the generated pseudo-code
may fail to preserve the original program semantics or support downstream reuse.

Recently, neural and LLM-based decompilation approaches (e.g., Dire~\cite{lacomis2019DIRENeural}, 
DeGPT~\cite{hu2024DeGPTOptimizing}, LLM4Decompile~\cite{tan2024LLM4Decompile}) generate readable code 
by learning syntactic patterns from large-scale pre-training. 
However, these methods primarily rely on statistical correlations between assembly and source code and optimize for textual similarity metrics such as BLEU~\cite{papineni2002bleu} and ROUGE~\cite{lin2004rouge}, rather than explicitly enforcing semantic correctness or executability. 
As a result, they often fail to 
account for compiler optimization semantics, leading to errors in types, boundary conditions, 
and control logic, especially for optimized binaries.

These limitations highlight two key challenges in executable decompilation.
\textbf{Challenge 1: Recovering Lost Semantics.} 
Stripped binaries lack explicit semantic
cues such as variable names and high-level control structures. Existing approaches struggle
to robustly infer this information, particularly under aggressive optimization.
\textbf{Challenge 2: Recompilability and Re-executability are second-class.} Even when decompiled code
appears plausible, it is rarely recompilable or behaviorally equivalent to the original
binary, limiting its practical usefulness in real-world scenarios.

To address these challenges, we present \textbf{ICL4Decomp}, an in-context learning
framework~\cite{dong2024SurveyIncontext,wies2023learnability} for executable decompilation. 
Rather than relying on model retraining or purely heuristic reconstruction, 
ICL4Decomp conditions a pretrained language model on carefully-designed contextual information at inference time. 
The framework integrates retrieved assembly--source exemplars and optimization-aware semantic guidance to improve both structural recovery and behavioral correctness of decompiled code.

We evaluate ICL4Decomp on multiple datasets across different compilers (GCC and Clang) 
and optimization levels (O0--O3), using re-executability rate as the primary metric, 
which measures whether the generated source code compiles and produces behavior 
consistent with the original binary on a held-out test suite. Experimental results 
demonstrate substantial improvements over state-of-the-art baselines, achieving an 
average increase of approximately 30\% in re-executability. The gains are particularly 
pronounced at higher optimization levels, where compiler transformations introduce 
greater semantic ambiguity, while ICL4Decomp maintains robust performance across all 
optimization settings.

Our main contributions are summarized as follows:
\begin{itemize}
    \item We introduce ICL4Decomp, an in-context learning framework that improves executable
    decompilation without model retraining.
    \item We demonstrate substantial gains in recompilability and semantic correctness
    across datasets, compilers, and optimization levels.
    \item We provide empirical analysis showing that contextual guidance improves robustness
    under aggressive compiler optimizations.
\end{itemize}


\section{Related Work}
\subsection{Neural Decompilation and Binary-to-Source Translation}
Traditional rule-based decompilers (e.g., Hex-Rays~\cite{IDAPro}, Ghidra~\cite{2025NationalSecurityAgencyGhidra}) rely on pattern heuristics that often fail under compiler optimizations and aggressive inlining.
Neural approaches instead learn correspondences between binary instructions and high-level abstractions.
Most methods follow the Neural Machine Translation (NMT) paradigm~\cite{sutskever2014sequence}.
Early work employs RNNs~\cite{katz2018UsingRecurrent} or combines NMT with program analysis (TraFix~\cite{katz2019NeuralDecompilation}).
Retargetable neural decompilation extends this paradigm across architectures~\cite{hosseini2022beyond}.
DIRE~\cite{lacomis2019DIRENeural} and Coda~\cite{fu2019coda} attempt to reconstruct source code but suffer from limited context windows and sparse vocabularies on complex ISAs.
Recent work, including LLM4Decompile~\cite{tan2024LLM4Decompile, tan2025DecompileBenchMillionScale}, SLADE~\cite{armengol2024slade},  and DecompileBench~\cite{gao2025DecompileBench}, leverages pretrained LLMs to achieve improved fluency.

However, current neural decompilers still struggle with cross-compiler generalization and semantic consistency~\cite{cao2023revisiting, kim2023revisiting}.
Reconstruction of high-level structures, such as loops and call hierarchies, remains limited, reflecting persistent difficulties in maintaining readability and structural coherence~\cite{vitale2025personalized, 10336303, sergeyuk2024reassessing}.
These limitations motivate the integration of retrieval-based and compiler-aware contextualization strategies.

\subsection{LLMs for Binary Analysis and Reasoning}
While large language models have redefined code understanding~\cite{chen2021evaluating,donato2025studying}, their application to binary analysis presents unique challenges.
General-purpose large language models can assist in reasoning over disassembly~\cite{jin2023binary}.
Concurrent research on transformer-based binary embeddings has enhanced control-flow representations~\cite{zhu2023ktrans}.
Frameworks such as ReSym~\cite{xie2024resym} demonstrate the feasibility of combining symbolic reasoning with pretrained code models.
Furthermore, source-code foundation models have been explored as transferable knowledge bases.
These studies show that code-pretrained large language models capture low-level semantics applicable to disassembled programs~\cite{su2024foundation}.

Despite this progress, empirical evaluations indicate that performance varies substantially across compilers and optimization levels~\cite{jin2023binary,shang2025binmetric}.
Studies document frequent hallucinations in code-oriented large language models that lead to functional errors~\cite{liu2024exploring}.
For stripped binaries, the lack of explicit compiler semantics degrades type and variable recovery unless augmented with external program analysis signals~\cite{xie2024resym,su2024foundation}.
These findings suggest that pure generation is insufficient.
Hybrid pipelines that pair large language models with external grounding are necessary to stabilize low-level reasoning.

\subsection{In-Context Learning for Program Synthesis}
In-context learning facilitates rapid adaptation to unseen codebases and problem styles by conditioning generation on exemplar demonstrations~\cite{brown2020language,nijkamp2023codegen2}.
Frameworks such as Self-refine and InCoder use few-shot exemplars to preserve syntactic correctness during code modification~\cite{madaan2023selfrefine,fried2022incoder}.
To enhance reliability, recent retrieval-augmented methods employ semantically similar exemplars drawn from large corpora to improve domain transfer~\cite{yang2025empirical}.

In the context of binary-to-source translation, in-context learning provides a natural mechanism to incorporate compiler- and optimization-specific context by retrieving representative assembly and source pairs~\cite{jin2023binary,su2024foundation}.
Empirical analyses show that exemplar similarity, measured via code embeddings or control-flow distance, strongly affects output fidelity~\cite{nijkamp2023codegen2,yang2025empirical}.
Integrating retrieval with in-context prompting thus offers a robust paradigm for cross-optimization decompilation.
This paradigm combines explicit structural grounding with the generalization power of pretrained large language models~\cite{shang2025binmetric}.

\section{Approach}
\label{sec:method}

We propose ICL4Decomp, an in-context learning framework for binary decompilation that aims to recover high-level, recompilable source code from optimized assembly. The key insight is that large language models can be guided to better reconstruct program structure and semantics when conditioned on carefully designed contextual information at inference time, without any model retraining.

\begin{figure}[t]
    \centering
    \includegraphics[width=1\linewidth]{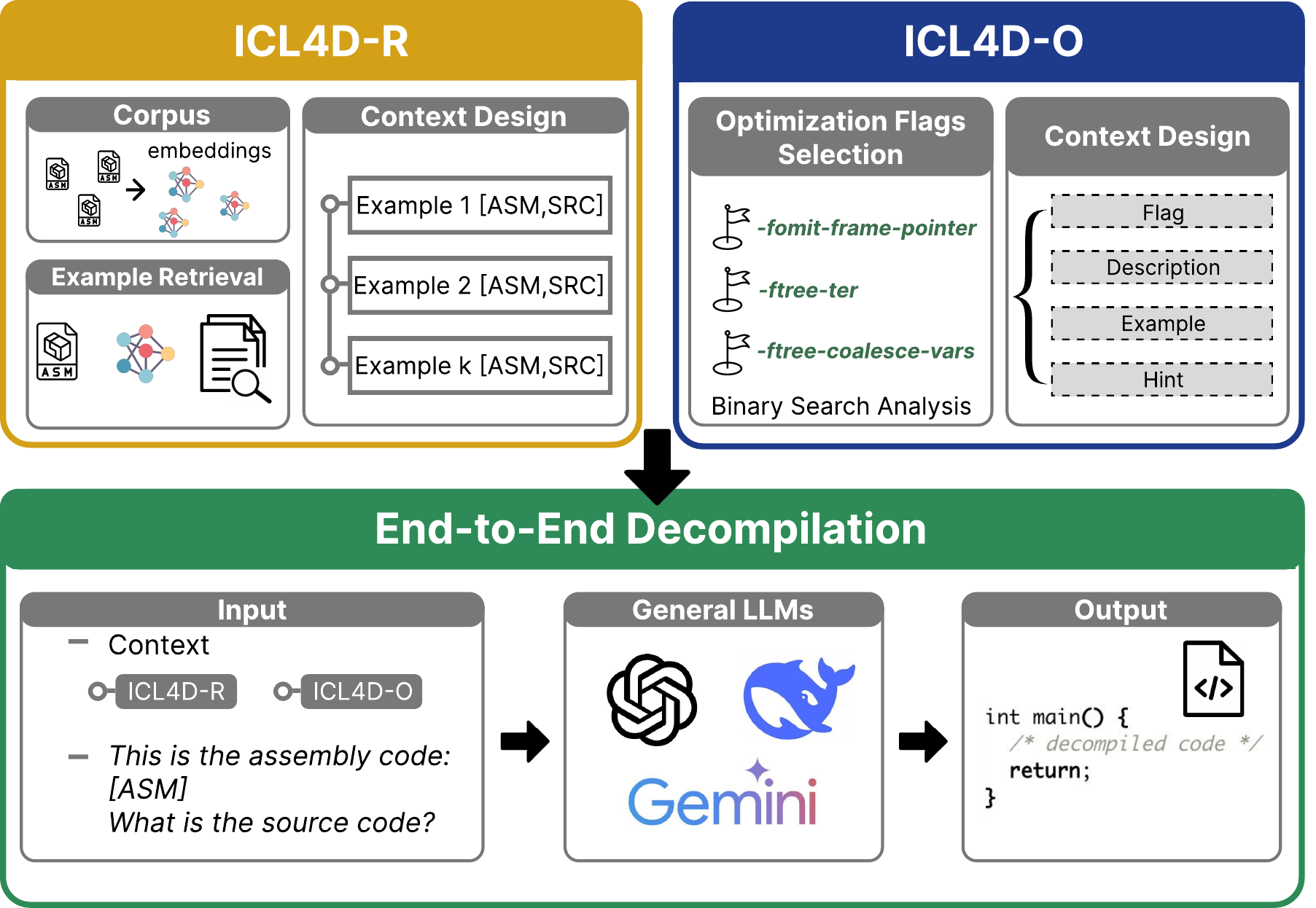}
    \caption{System overview for in-context decompilation.}
    \label{fig:retrieval_pipeline}
\end{figure}

Given a target assembly function, ICL4Decomp constructs an informative context and conditions a frozen large language model to directly generate corresponding source code. The framework incorporates two complementary forms of contextual guidance: (i) retrieved assembly–source exemplars that expose concrete instruction-to-structure correspondences, and (ii) optimization-aware natural language rules that encode compiler transformation semantics. \autoref{fig:retrieval_pipeline} provides an overview of the framework.

\subsection{Problem Setup}
Let $A$ denote a compiled assembly function and $S$ its corresponding high-level source implementation. The goal of decompilation is to recover a source program $\hat{S}$ such that, when recompiled, it exhibits behavior equivalent to the original binary. In this work, we focus on \emph{re-executable decompilation}, where success requires both syntactic correctness (the generated code compiles) and semantic equivalence under a held-out test suite.

We formalize in-context decompilation as conditional generation with a frozen language model $f_\theta$:
\[
\hat{S} = f_\theta(A, C),
\]
where $C$ denotes auxiliary contextual information provided at inference time. No parameter updates or finetuning are performed; all adaptation occurs purely through prompt conditioning.

\subsection{Retrieved-Exemplar In-Context Decompilation (ICL4D-R)}
\label{sec:icl4dr}

The first variant of our framework, \textbf{ICL4D-R}, leverages retrieved assembly--source exemplars as in-context demonstrations. Given a target assembly function $A_t$, we retrieve a small set of semantically similar function pairs $\{(A_i, S_i)\}$ from a preconstructed corpus and include them in the prompt prior to the target assembly.

Each exemplar provides an explicit mapping between low-level instructions and high-level program structure, allowing the model to observe how control flow, expressions, and variable usage are recovered under similar compilation patterns. By conditioning on multiple such demonstrations, the language model implicitly adapts to the compiler style and optimization level of the target function.

The retrieved exemplars are ordered by semantic similarity and formatted as alternating assembly and source segments, followed by the target assembly and a decompilation instruction. This structured prompting exposes the model to concrete instruction-to-structure correspondences before generation. 
Details of corpus construction, embedding, similarity computation, and retrieval are provided in Appendix~\ref{appendix:retrieval}, 
while the prompt formatting and exemplar organization used for in-context decompilation are described in Appendix~\ref{appendix:prompts}.

\subsection{Optimization-Aware In-Context Decompilation (ICL4D-O)}
\label{sec:icl4do}

While retrieved exemplars are effective when local instruction--structure correspondences are preserved, aggressive compiler optimizations often introduce non-local transformations that obscure data flow and control structure. To address this challenge, we introduce \textbf{ICL4D-O}, a rule-based variant that augments the prompt with optimization-aware contextual guidance.

ICL4D-O encodes compiler optimization semantics as natural language descriptions that explain how specific transformations affect the emitted assembly. These rules inform the model that certain source-level constructs, such as temporary variables or stack frames, may be absent due to semantics-preserving compiler optimizations, and should be reconstructed accordingly during decompilation.

Rather than attempting to infer optimization behavior implicitly, ICL4D-O explicitly conditions generation on relevant optimization rules, guiding the model's reasoning about data flow, variable lifetimes, and control structure. This approach is particularly beneficial at higher optimization levels, where conventional decompilation and purely exemplar-based prompting tend to fail. The identification of influential optimization flags and the design of rule-based prompts are described in Appendix~\ref{appendix:flags} and Appendix~\ref{appendix:prompts}, respectively.

\subsection{End-to-End Decompilation Pipeline}
\label{sec:end_to_end}

Both ICL4D-R and ICL4D-O operate in an end-to-end manner at inference time. Given a target assembly function, the framework constructs a prompt by selecting either retrieved exemplars, optimization-aware rules, or both, and conditions a pretrained language model to generate the corresponding source code in a single forward pass. No symbolic execution, control-flow reconstruction, or post-hoc repair is performed.

This design combines the flexibility of in-context learning with compiler-aware semantic grounding, enabling the generation of source code that is not only readable but also recompilable and executable across diverse optimization levels.

\section{Evaluation}
\label{sec:evaluation}
\begin{table*}[!htbp]
\caption{Re-executability rate (\%) comparison on HumanEval and ExeBench datasets. \textbf{Ghidra} represents the traditional decompilation baseline.}
\label{tab:res-examples}
\centering
\begin{tabularx}{\textwidth}{lYYYYY|YYYYY}
\toprule
\multirow{2}{*}{\textbf{Method}} 
& \multicolumn{5}{c|}{\textbf{HumanEval-Decompile}} 
& \multicolumn{5}{c}{\textbf{ExeBench}} \\
\cmidrule{2-11}
 & O0 & O1 & O2 & O3 & AVG & O0 & O1 & O2 & O3 & AVG \\ 
\midrule
Ghidra 
& 12.50 & 17.07 & 12.50 & 11.28 & 13.34 
& 16.37 & 18.94 & 16.66 & 17.64 & 17.40 \\ 
\note{}IDA Pro 
&  & 17.07 & 12.50 & 11.28 & 13.34 
& 16.37 & 18.94 & 16.66 & 17.64 & 17.40 \\ 
\midrule
LLM4Decompile-1.3B 
& 26.78 & 11.53 & 13.22 & 11.53 & 15.77
& 15.10 & 12.83 & 13.22 & 11.53 & 13.17 \\
DeepSeek-V3.2 
& 46.65 & 32.32 & 33.23 & 35.06 & 36.82
& 26.17 & 19.26 & 19.79 & 16.13 & 20.34 \\ 
\midrule
\textbf{ICL4D-R} 
& \textbf{54.27} & \textbf{42.38} & \textbf{40.24} & \textbf{42.07} & \textbf{44.74}
& \textbf{34.39} & \textbf{35.68} & \textbf{36.16} & \textbf{33.74} & \textbf{34.99} \\
\bottomrule
\end{tabularx}
\end{table*}

In this section, we evaluate our in-context decompilation framework by addressing three research questions:
\begin{itemize}
    \item \textbf{RQ1 (Executability):} Can in-context learning improve the re-executable rate compared to baselines?
    \item \textbf{RQ2 (Error Mitigation):} Does in-context guidance mitigate specific compilation and runtime errors?
    \item \textbf{RQ3 (Robustness):} How robust is the framework across varying program complexities?
\end{itemize}

\subsection{Experimental Setup}
\label{sec:setup}

\paragraph{Datasets.}
We utilize two datasets covering system-level and algorithmic domains. Statistics are detailed in Table~\ref{tab:dataset-stats}.
\textsc{(i) ExeBench}~\cite{armengol-estape2022ExeBenchMLscalea}: A machine-learning-scale 
dataset of real-world C functions. We use the \texttt{test-real} subset ($N=1,135$), containing functions with concrete auxiliary definitions suitable for I/O-driven synthesis.
\textsc{(ii) HumanEval-Decompile}: Adapted from \textsc{HumanEval}~\cite{chen2021evaluating}, comprising 1,312 algorithmic Python problems translated into executable C/C++ by LLM4Decompile~\cite{tan2024LLM4Decompile}.
Both datasets cover optimization levels O0--O3. All samples are normalized (see \S3.2) and validated by SHA-256 to ensure no overlap with the retrieval corpus.

\begin{table}[!htbp]
\centering
\setlength{\tabcolsep}{3pt}
\renewcommand{\arraystretch}{1.05}
\caption{Evaluation dataset statistics (mean / std) per function.}
\label{tab:dataset-stats}
\resizebox{\columnwidth}{!}{%
\begin{tabular}{l r r r r}
\toprule
\textbf{Evaluation Dataset} & \textbf{\#N} & \textbf{LOC} & \textbf{Cycl.} & \textbf{Blocks}\\
\midrule
ExeBench & 1135 &
\makecell{\phantom{$\pm$}13.6\\$\pm$16.1} &
\makecell{\phantom{$\pm$0}3.4\\$\pm$15.3} &
\makecell{\phantom{$\pm$0}3.5\\$\pm$15.3} \\
HumanEval-Decompile & 1312 &
\makecell{\phantom{$\pm$}13.3\\\phantom{0}$\pm$8.4} &
\makecell{\phantom{$\pm$0}4.9\\\phantom{0}$\pm$3.1} &
\makecell{\phantom{$\pm$0}5.1\\\phantom{0}$\pm$3.0} \\
\bottomrule
\end{tabular}%
}
\end{table}

\paragraph{Implementation \& Baselines.}
We employ \textsc{DeepSeek-V3.2} as the primary generation model and \textsc{Nova-1.3b}~\cite{jiang2023nova} as the embedding encoder (1024-d).
We compare our approach against:
(1) \textbf{DeepSeek-V3.2} (Zero-shot baseline);
(2) \textbf{LLM4Decompile-End}~\cite{tan2024LLM4Decompile} (SOTA learning-based baseline);
(3) \textbf{Ghidra} (Traditional rule-based baseline.
We evaluate two variants of our framework: \textbf{ICL4D-R} (Retrieval-Augmented) and \textbf{ICL4D-O} (Rule-Guided).
Generation uses temperature $0.1$, exemplar count $k=5$, and max tokens $10{,}000$. 

\paragraph{Metrics.}
We report the \textbf{Executable Success Rate (ESR)}, defined as the proportion of decompiled functions $\hat{S}_t$ that compile at \texttt{-O0} and pass all I/O test cases within a 5-second timeout, ensuring both syntactic and semantic correctness.

\subsection{RQ1: Improvement in Executability}
\label{sec:rq1}

We first assess the overall executability improvements provided by our in-context strategies.

\paragraph{Performance of ICL4D-R.}
As shown in Table~\ref{tab:res-examples}, \textbf{ICL4D-R} consistently outperforms all baselines across both datasets and all optimization levels.
On \textit{HumanEval-Decompile}, it achieves \textbf{54.3\%} accuracy at O0, surpassing DeepSeek-V3.2 (46.7\%) and LLM4Decompile (26.8\%).
On the more complex \textit{ExeBench}, ICL4D-R demonstrates up to a \textbf{30\% improvement} over learning-based baselines.
This confirms that retrieval-based exemplars effectively help the model generalize across diverse compiler transformations.

\paragraph{Performance of ICL4D-O.}
Since decompilation failures increase at higher optimization levels, we apply the rule-based \textbf{ICL4D-O} specifically to recover samples that failed in the first round (O1--O3).
Table~\ref{tab:reexe_rules_dual} shows that while less stable than retrieval, ICL4D-O yields selective gains. For instance, prompting with \texttt{-ftree-coalesce-vars} at O2 improves executability to 17.35\% on HumanEval.
While rule-based prompting can be rigid, manual inspection suggests it produces more localized, repairable errors, which we analyze further in RQ2.

\begin{table}[!htbp]
\centering
\caption{Re-executability rate (\%) after applying rule-based prompts (ICL4D-O) for different optimization options across two datasets. Values in parentheses indicate baseline performance.}
\label{tab:reexe_rules_dual}
\resizebox{\columnwidth}{!}{%
\footnotesize
\begin{tabular}{lrr}
\toprule
\textbf{O1 Option} & \textbf{ExeBench (11.18)} & \textbf{HumanEval (14.86)} \\
\midrule
\texttt{-fomit-frame-pointer}   & 9.34 & \textbf{15.32} \\
\texttt{-ftree-ter}             & 8.17 & \textbf{15.77} \\
\texttt{-ftree-coalesce-vars}   & \textbf{11.67} & 12.61 \\
\texttt{-fipa-pure-count}       & 8.95 & 11.26 \\
\midrule
\textbf{O2 Option} & \textbf{ExeBench (7.48)} & \textbf{HumanEval (15.98)} \\
\midrule
\texttt{-fomit-frame-pointer}   & \textbf{9.52}  & 12.41 \\
\texttt{-ftree-ter}             & \textbf{7.87}  & \textbf{16.89} \\
\texttt{-ftree-coalesce-vars}   & \textbf{10.63} & \textbf{17.35} \\
\texttt{-fipa-pure-count}       & 6.72  & 10.50 \\
\texttt{-fcrossjumping}         & \textbf{7.91}  & 5.05 \\
\midrule
\textbf{O3 Option} & \textbf{ExeBench (3.18)} & \textbf{HumanEval (15.02)} \\
\midrule
\texttt{-fomit-frame-pointer}   & \textbf{8.18}  & \textbf{14.08} \\
\texttt{-ftree-ter}             & \textbf{5.48}  & \textbf{15.49} \\
\texttt{-ftree-coalesce-vars}   & \textbf{10.96} & 11.27 \\
\texttt{-fipa-pure-count}       & \textbf{6.40}  & 7.51 \\
\texttt{-fcrossjumping}         & \textbf{5.45}  & \textbf{12.21} \\
\bottomrule
\end{tabular}%
}
\end{table}

\begin{tcolorbox}[colback=gray!5,colframe=gray!40!black,boxrule=0.4pt,boxsep=-1mm]
\textbf{Answer to RQ1:} In-context learning dramatically improves executability. \textbf{ICL4D-R} provides robust gains across all settings, while \textbf{ICL4D-O} offers targeted improvements for high-optimization failures.
\end{tcolorbox}

\subsection{RQ2: Error Analysis and Mitigation}
\label{sec:rq2}

To understand \textit{how} in-context learning improves performance, we analyze the shift in failure modes using a taxonomy derived from compiler \texttt{stderr} diagnostics (Table~\ref{tab:error-taxonomy}).

\begin{table}[!htbp]
\caption{Error taxonomy for \texttt{stderr} diagnostics.}
\label{tab:error-taxonomy}
\setlength{\tabcolsep}{3pt}
\renewcommand{\arraystretch}{0.95}
\footnotesize
\begin{tabularx}{\columnwidth}{l X}
\toprule
\textbf{Category} & \textbf{Example cause / message} \\
\midrule
Assert & Assertion failure or output mismatch (\textit{assertion failed}) \\
Syntax & Token or structure error (\textit{expected ';'}, \textit{unterminated string}) \\
Return & Invalid return or argument count (\textit{void value not ignored}) \\
Type & Incompatible or invalid type (\textit{invalid conversion}) \\
Declaration & Missing or conflicting symbol (\textit{undefined reference}) \\
Runtime/Link & Crash or linking error (\textit{segmentation fault}) \\
Other & Uncategorized message \\
\bottomrule
\end{tabularx}
\vspace{-2mm}
\end{table}

\begin{figure}[!htbp]
    \centering
    \begin{subfigure}[b]{0.48\linewidth}
        \centering
        \includegraphics[width=\linewidth]{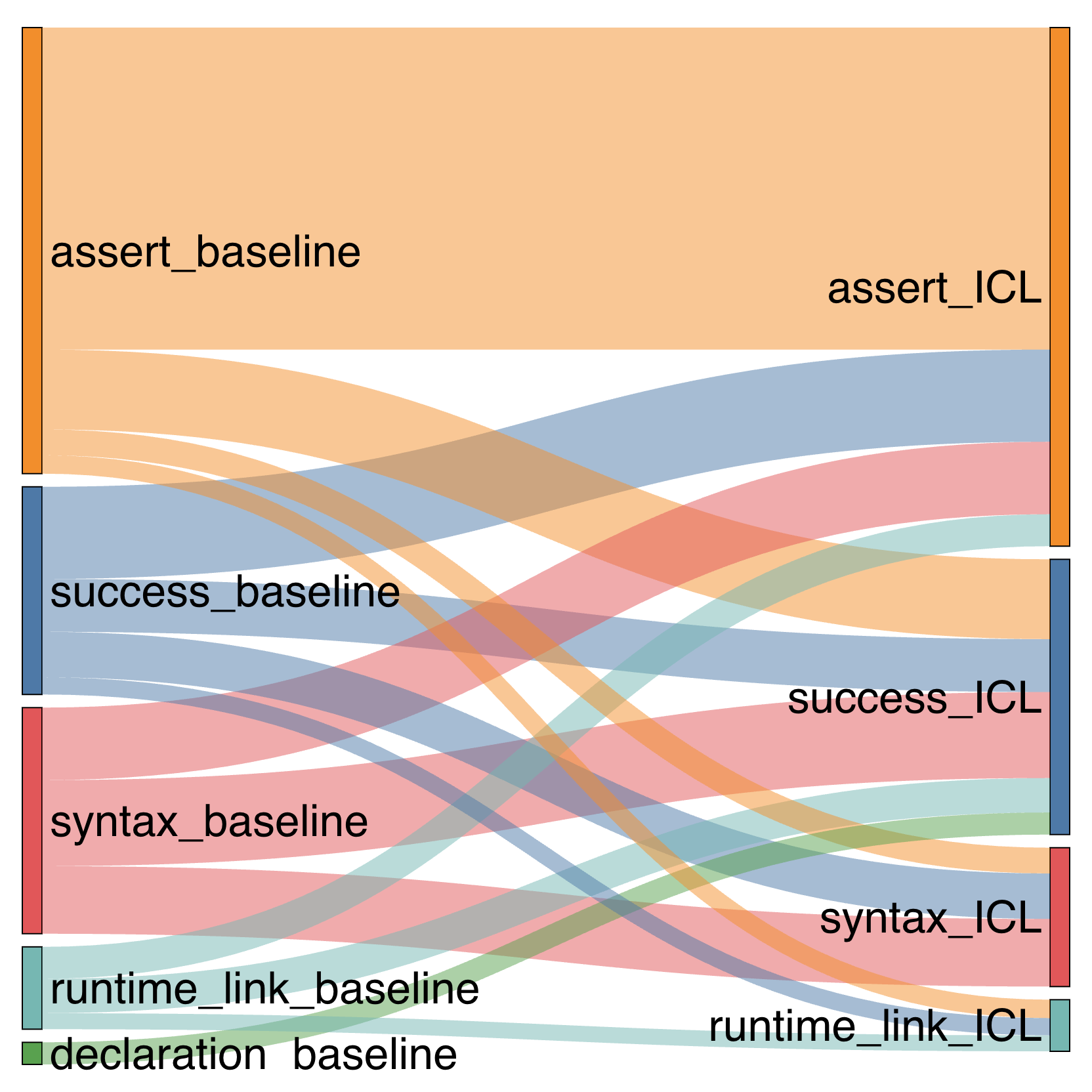}
        \caption{HumanEval dataset.}
        \label{fig:sankey_errorflow_a}
    \end{subfigure}
    \hfill
    \begin{subfigure}[b]{0.48\linewidth}
        \centering
        \includegraphics[width=\linewidth]{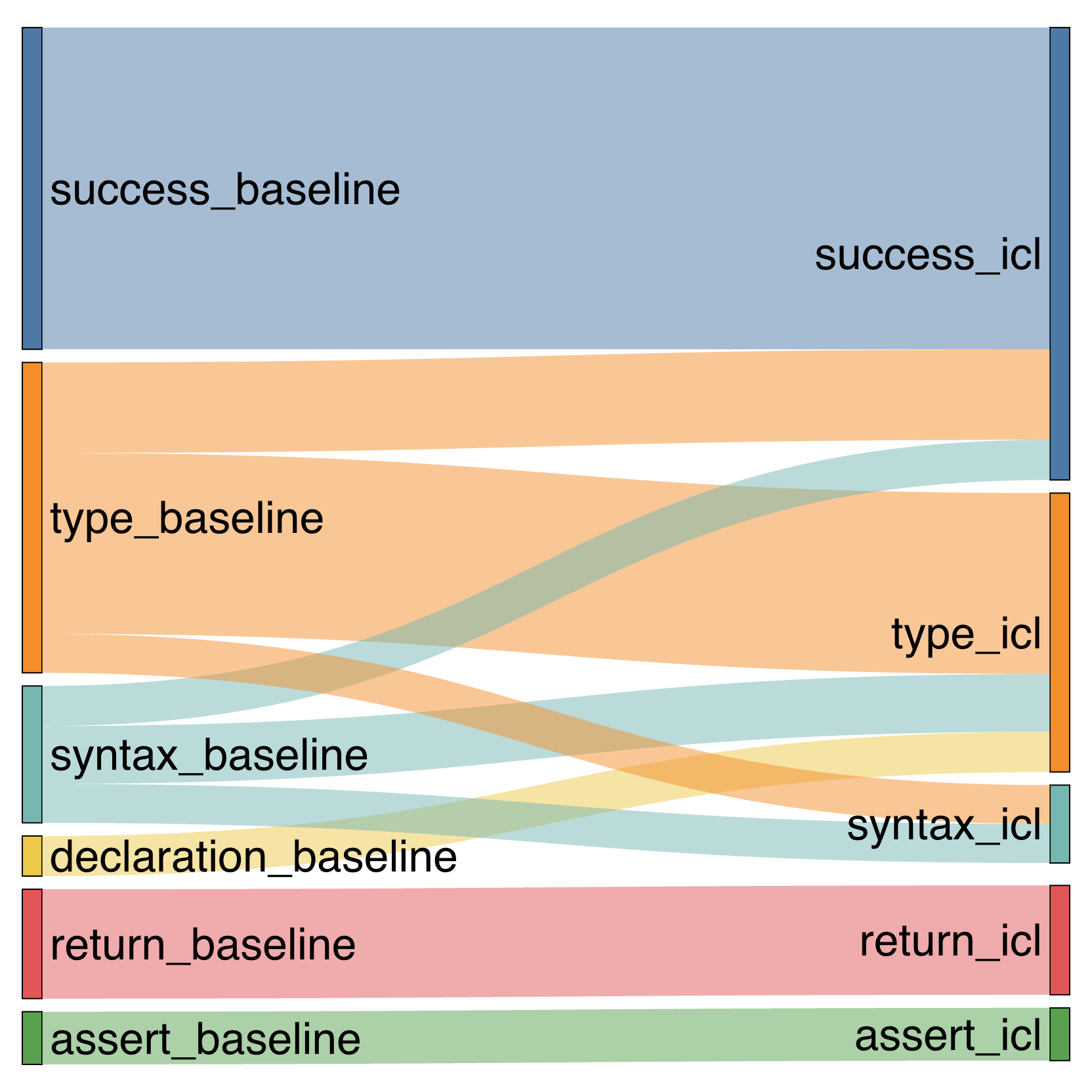}
        \caption{ExeBench dataset.}
        \label{fig:sankey_errorflow_b}
    \end{subfigure}

    \caption{Distribution shift of error categories before and after applying in-context learning. }
    \label{fig:sankey_errorflow}
\end{figure}

\paragraph{Error Distribution Shift.}
Figure~\ref{fig:sankey_errorflow} illustrates the transition of error categories from the baseline to ICL4D-R.
\begin{itemize}
    \item \textbf{HumanEval (Fig.~\ref{fig:sankey_errorflow_a}):} We observe a substantial reduction in \textit{Syntax}, \textit{Runtime}, and \textit{Declaration} errors converting into successes. This indicates that exemplars aid in reconstructing valid syntactic structures and resolving undefined symbols.
    \item \textbf{ExeBench (Fig.~\ref{fig:sankey_errorflow_b}):} Improvements are concentrated in \textit{Type} and \textit{Syntax} categories. Given ExeBench's structural complexity, this suggests ICL effectively enhances semantic reasoning and type inference.
\end{itemize}

\paragraph{Rule-Guided Error Localization.}
For ICL4D-O, Table~\ref{tab:humaneval_rulewise} reveals a trade-off: while it reduces \textit{Declaration} and \textit{Type} errors, it often increases \textit{Syntax} errors.
This shift is beneficial: global structural failures (like missing declarations) are converted into localized syntax errors, which are generally easier to repair.

\begin{table}[t]
\centering
\caption{HumanEval (ICL4D-O) rule-wise error distribution (\%), aggregated over O1--O3.
Baseline is averaged over O1--O3. Arrows ($\downarrow$/$\uparrow$/$\approx$) indicate change relative to baseline.
Rule abbreviations: FC (\texttt{-ftree-coalesce-vars}), FP (\texttt{-fipa-pure-const}),
FT (\texttt{-ftree-ter}), FFP (\texttt{-fomit-frame-pointer}), FJ (\texttt{-fcrossjumping}).}
\label{tab:humaneval_rulewise}

\small
\setlength{\tabcolsep}{3.5pt}
\begin{tabular}{lrrrrrr}
\toprule
\textbf{Category} & \textbf{Base} & \textbf{FC} & \textbf{FP} & \textbf{FT} & \textbf{FFP} & \textbf{FJ} \\
\midrule
Syntax      & 15.3 & 17.2$\uparrow$ & 15.4$\approx$ & 19.8$\uparrow$ & 16.0$\uparrow$ & 25.8$\uparrow$ \\
Declaration & 27.3 & 12.3$\downarrow$ & 30.9$\uparrow$ & 23.3$\downarrow$ & 23.9$\downarrow$ & 16.4$\downarrow$ \\
Type        &  6.0 &  4.6$\downarrow$ &  8.8$\uparrow$ &  8.1$\uparrow$ &  5.6$\approx$ &  8.3$\uparrow$ \\
Return      &  0.7 &  1.1$\approx$    &  0.2$\downarrow$ &  1.6$\uparrow$ &  0.7$\approx$ &  0.5$\approx$ \\
Assert      & 41.5 & 49.6$\uparrow$ & 40.3$\downarrow$ & 39.0$\downarrow$ & 42.9$\uparrow$ & 33.3$\downarrow$ \\
Other       &  9.1 & 15.3$\uparrow$ &  4.4$\downarrow$ &  8.1$\downarrow$ & 10.8$\uparrow$ & 15.6$\uparrow$ \\
\bottomrule
\end{tabular}
\end{table}

\begin{figure}[!htbp]
    \centering
    \includegraphics[width=1\linewidth]{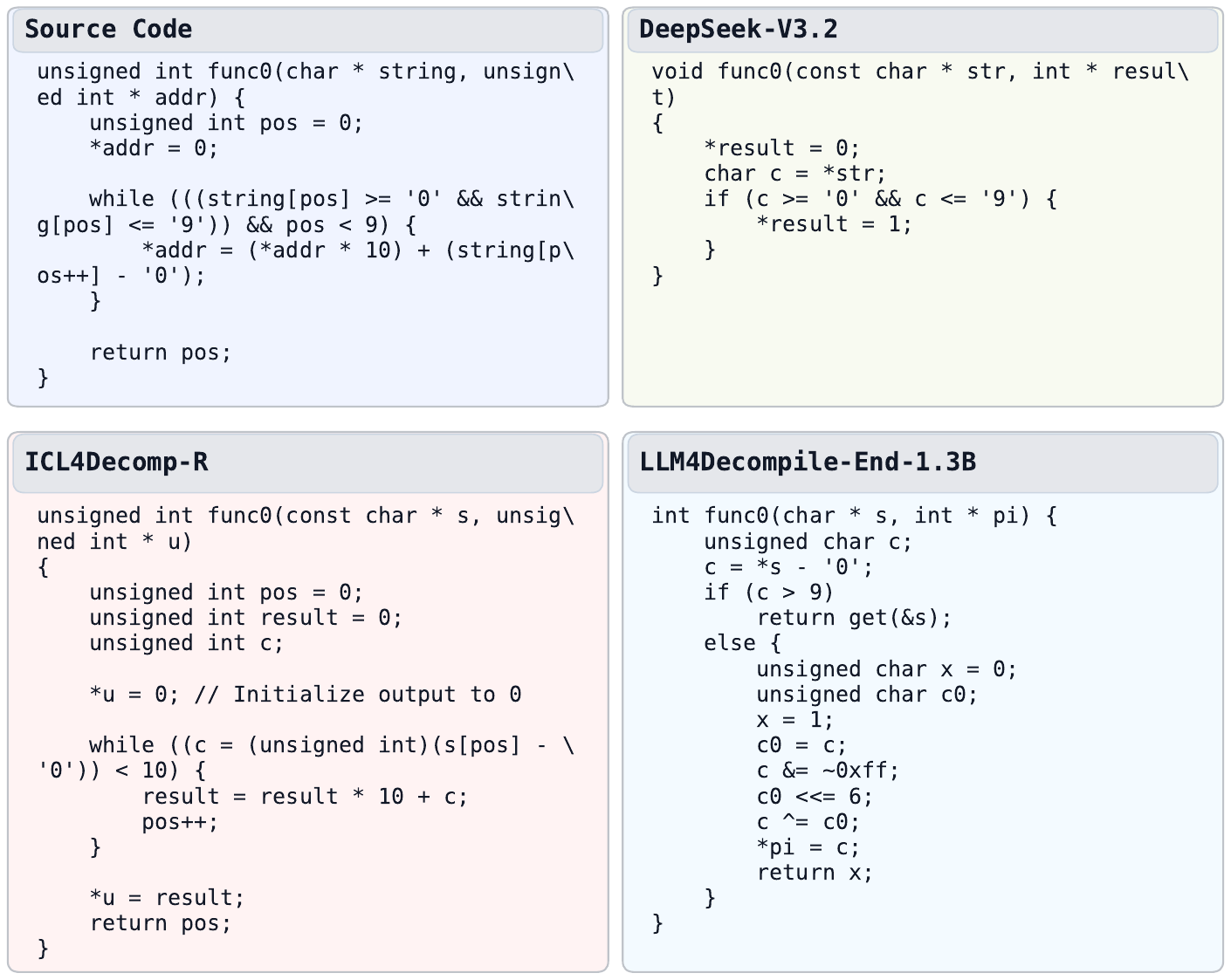}
    \caption{Qualitative example: Ground-truth vs. decompilations from three methods.}
    \label{fig:error_example_exebench}
\end{figure}

\paragraph{Qualitative Analysis.}
Figure~\ref{fig:error_example_exebench} presents a case study from ExeBench.
The ground truth involves parsing digits into an integer.
\textbf{DeepSeek-V3.2} hallucinates a constant boolean return, and \textbf{LLM4Decompile} introduces spurious bitwise operations.
In contrast, \textbf{ICL4D-R} correctly recovers the loop bounds and arithmetic logic, demonstrating that retrieval contexts prevent control-flow collapse and semantic hallucination.

\begin{tcolorbox}[colback=gray!5,colframe=gray!40!black,boxrule=0.4pt,boxsep=-1mm]
\textbf{Answer to RQ2.} 
In-context learning mitigates structural errors in decompilation. 
\textbf{ICL4D-R} effectively reduces syntax and declaration failures, improving structural and symbolic consistency, 
while \textbf{ICL4D-O} produces more localized and repairable errors.
\end{tcolorbox}

\subsection{RQ3: Robustness and Ablation}
\label{sec:rq3}

\begin{figure}[!htbp]
    \centering
    \includegraphics[width=1\linewidth]{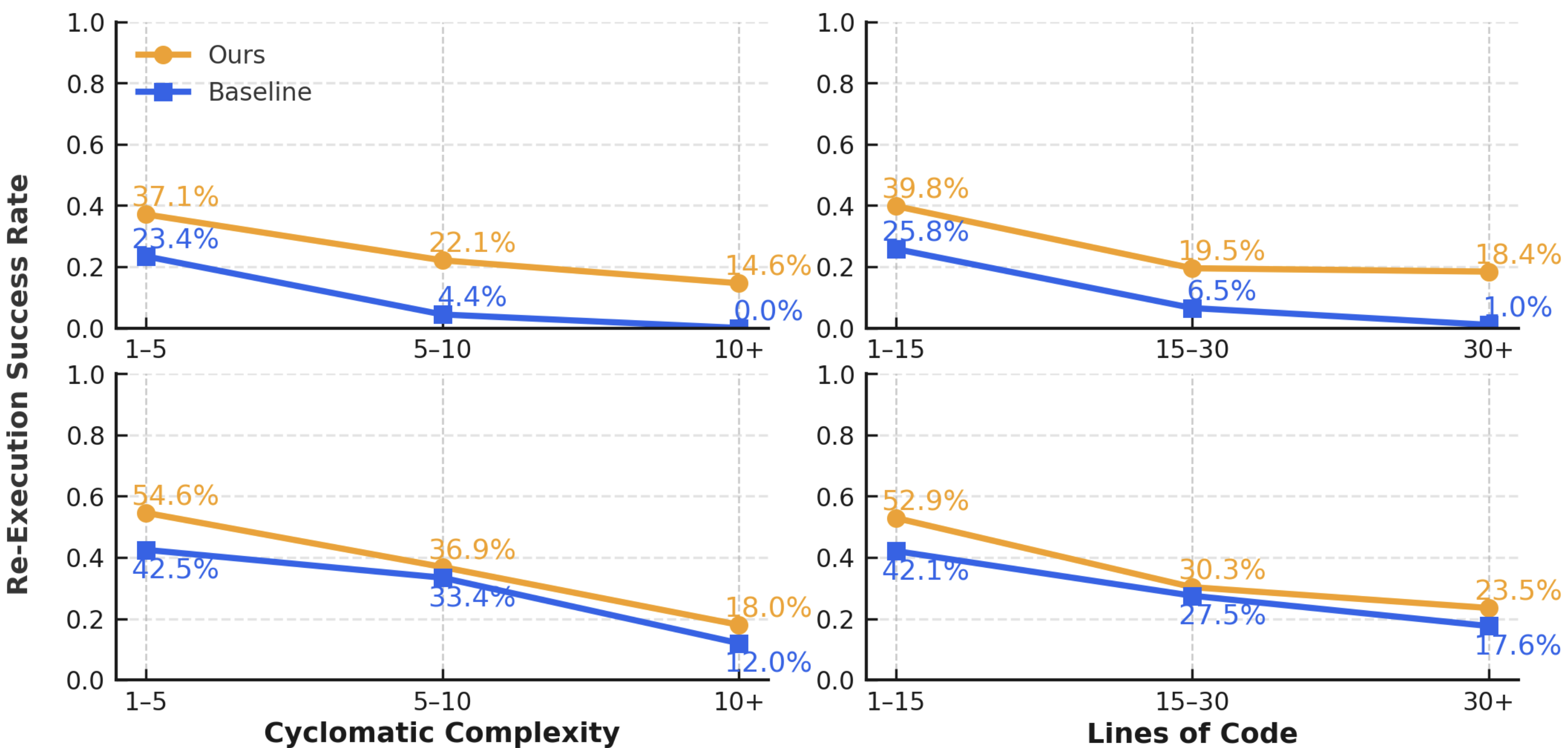}
    \caption{Re-execution success rate across functions of varying cyclomatic complexity and lines of code for HumanEval-Decompile (top) and ExeBench (bottom).}
    \label{fig:robustness}
\end{figure}

\paragraph{Robustness (RQ3).}
We stratify performance by Cyclomatic Complexity and LOC (Figure~\ref{fig:robustness}).
While all models degrade as complexity increases, ICL4D-R (Ours) exhibits a slower rate of decay, particularly in the mid-complexity range (5–10 branches). This indicates stronger generalization to intricate control flows compared to the baseline.

\begin{tcolorbox}[colback=gray!5,colframe=gray!40!black,boxrule=0.4pt,boxsep=-1mm]
\textbf{Answer to RQ3.}
ICL4D-R demonstrates strong robustness across program complexity, consistently outperforming the baseline and maintaining higher re-execution success rates even as cyclomatic complexity and code length increase.
\end{tcolorbox}

\paragraph{Ablation Study.}

To disentangle the effect of \textit{semantic retrieval} from that of merely providing \textit{additional few-shot context}, 
we construct a controlled baseline termed \textbf{Random Retrieval} (Table~\ref{tab:ablation_random_retrieval}). 
This variant preserves the \emph{exact prompt structure, exemplar count ($k=5$), and context length}, 
but replaces semantically matched assembly–source pairs with randomly sampled ones.

Performance degrades substantially and consistently across datasets and optimization levels. 
For example, on HumanEval-Decompile at O3, executability drops from 42.07\% to 0.35\%. 
Similar degradation is observed on ExeBench (e.g., O3: 33.74\% → 15.72\%). 

These results demonstrate that improvements are not attributable to extended context alone. 
Instead, semantic alignment of retrieved exemplars is essential for guiding structural and behavioral reconstruction during decompilation.

\begin{table}[!htbp]
\caption{Ablation on random retrieval control. The random-retrieval variant replaces semantically similar examples with random samples while preserving context length.}
\label{tab:ablation_random_retrieval}
\centering
\begin{tabularx}{0.48\textwidth}{lYYYY}
\toprule
\multicolumn{5}{c}{\textbf{HumanEval-Decompile}} \\
\midrule
Model & O0 & O1 & O2 & O3 \\ \midrule
DeepSeek-V3.2    & 46.65 & 32.32 & 33.23 & 35.06 \\
\textbf{ICL4D-R} & \textbf{54.27} & \textbf{42.38} & \textbf{40.24} & \textbf{42.07} \\
ICL4D-R (Ablation) & 42.25 & 40.49 & 41.34 & 0.35 \\
\midrule
\multicolumn{5}{c}{\textbf{ExeBench}} \\
\midrule
Model & O0 & O1 & O2 & O3 \\ \hline
DeepSeek-V3.2    & 26.17 & 19.26 & 19.79 & 16.13 \\
\textbf{ICL4D-R} & \textbf{34.39} & \textbf{35.68} & \textbf{36.16} & \textbf{33.74} \\
ICL4D-R (Ablation) & 25.50 & 21.62 & 19.45 & 15.72 \\
\bottomrule
\end{tabularx}
\end{table}

\section{Discussion}
\subsection{Applicability}
ICL4Decomp is designed to be model-agnostic as ICL can be readily adapted to many existing LLM architecture,
including open-source models such as \textit{DeepSeek}, \textit{Qwen}, and \textit{CodeLlama}, as well as commercial models like \textit{GPT-4} and \textit{Claude}. 
For scenarios involving privacy or code security concerns, ICL4Decomp can be deployed locally with open-source models to enable offline decompilation. 
The framework is fully decoupled at the API layer, making it facilitating integration into existing analysis platforms like IDA or Ghidra through plugin interfaces. 
Because ICL4Decomp operates entirely at inference time and does not require any model retraining, it is particularly suitable for environments with limited resources or cases where the original source code is unavailable.  
In practice, our framework can be applied to several binary analysis tasks, including security auditing, patch analysis, and reverse engineering, while maintaining consistent performance across different compiler optimization levels.

\subsection{Cost Analysis}
The primary cost of ICL4Decomp arises from API-based LLM invocations, which include:
(a) the assembly code and retrieved exemplars used to construct the in-context prompt, and 
(b) the generated source code output. 
Since individual functions are relatively short (typically under 300 tokens), the per-sample inference cost remains significantly lower than in general-purpose code generation tasks. 
For example, when using \textit{DeepSeek V3.2} on the HumanEval-Decompile dataset, the total cost for decompiling the entire benchmark is approximately \$3.

In terms of runtime, ICL4Decomp achieves high throughput through a multi-worker parallel execution mechanism. 
By distributing decompilation requests across multiple workers on a single multi-core workstation, decompiling 1{,}000 functions requires only about 15 minutes. 
This near-linear scalability with respect to the number of workers makes the framework suitable for large-scale binary analysis in real-world settings. 
Overall, ICL4Decomp demonstrates strong portability across models and deployment environments, and its inference cost scales linearly with workload size, supporting practical use in security and engineering applications.

\section{Conclusion}
This paper presents \textbf{ICL4Decomp}, a unified in-context learning framework for executable decompilation that jointly addresses syntactic recovery and semantic reasoning. 
It introduces two complementary mechanisms: \textbf{ICL4D-R}, which retrieves assembly–source exemplars to provide structural priors, and \textbf{ICL4D-O}, which injects optimization semantics through rule-based natural-language prompts. 
Together, they enable bidirectional alignment between syntax and semantics during inference.

Across ExeBench and HumanEval-Decompile, ICL4Decomp improves the re-executability rate by approximately 40\% on average, with particularly strong gains under higher optimization levels (O1–O3). 
It effectively reduces syntax and declaration errors, while maintaining robustness to function complexity and context length. 
These results demonstrate that in-context learning offers a practical and reproducible way to achieve executable consistency without model retraining.

Future work will explore hybrid context composition between retrieval and rule-based prompting, automatic inference of compiler optimization patterns, and broader generalization across compilers and architectures. 
Overall, ICL4Decomp reveals the potential of in-context learning to advance decompilation from \textit{syntactic readability} toward \textit{semantic executability}.

\section*{Limitations}
While our framework demonstrates promising results, we acknowledge several limitations inherent to our current methodology and evaluation scope.

First, our evaluation is scoped to the function level, prioritizing the syntactic and semantic re-executability of binaries obtained from isolated code units. 
We leverage two representative benchmarks, ExeBench and HumanEval-Decompile, reporting results stratified by structural complexity metrics such as lines of code, cyclomatic complexity, and basic block count (cf. \autoref{sec:evaluation}). 
Although system-level and multi-module programs are not the primary focus of this study, our framework is agnostic to dataset-specific features; extending it to broader codebases primarily requires adjusting the context composition strategy. 
Nonetheless, this protocol aligns with established best practices for rigorously evaluating LLM-based decompilation techniques~\cite{grubisic2024compiler,tan2024LLM4Decompile,dramko2025IdiomsNeurala,feng2025ReFDecompile}.

Second, while our experiments span optimization levels from O0 to O3, the dataset generation process is restricted to single-function compilation. 
Consequently, certain global optimization patterns (e.g., Link-Time Optimization) are excluded, and our results may be conservative when applied to scenarios involving aggressive cross-module optimizations.

Finally, 
the current implementation of ICL4D-O employs a predefined pool of optimization rules to construct contextual prompts. 
We intentionally avoid dynamically inferring the active optimization set from each assembly function, as mapping assembly patterns back to specific compiler flags is inherently ambiguous, many flags interact or produce overlapping instruction-level effects. 
However, enabling adaptive rule selection based on automatically detected optimization patterns could enhance contextual relevance and represent a promising avenue for future exploration.

\section*{Ethics Statement}
We recognize that better decompilation tools can carry risks. There is a possibility that this technology could be used to steal proprietary code or bypass software licensing. 
However, our primary goal is to help security analysts and developers who need to understand binary code when the original source is unavailable, such as analyzing malware or maintaining legacy systems. 
We aim to support these legitimate uses, not to facilitate copyright infringement or the theft of intellectual property.
Moreover, our framework is that it relies on in-context learning rather than model training. We only use publicly available datasets for retrieval and do not require access to private codebases or user data for fine-tuning.


\bibliography{custom}

\appendix

\section{Retrieval Infrastructure and Corpus Construction}
\label{appendix:retrieval}

This appendix provides implementation details for the retrieval-based in-context learning variant (ICL4D-R), including corpus construction, normalization, embedding, similarity computation, and the retrieval procedure.

\subsection{Retrieval Corpus Construction}

To support retrieval-based in-context decompilation, we construct a corpus of paired assembly--source functions $(A_i, S_i)$ drawn from two widely adopted datasets: MBPP (Mostly Basic Programming Problems) and ExeBench.

MBPP is commonly used to evaluate code generation models and provides algorithmic-level programs, while ExeBench contains system-level C functions collected from real-world GitHub projects. Combining these datasets yields a diverse corpus covering both algorithmic and system-oriented programming patterns. In total, the corpus contains approximately $10^5$ unique function pairs.

To ensure consistency across compilers and architectures, all assembly functions are normalized following the preprocessing used by the NOVA foundation model. Corresponding source code is also normalized to a canonical format. Each function is further annotated with a functional category (algorithm, string, I/O, system, or math), inferred from library headers and API usage. Exact duplicate pairs are removed via content hashing.

All retrieval corpus functions are strictly disjoint from the evaluation datasets (ExeBench test-real and HumanEval-Decompile). Disjointness is verified using SHA-256 hashing to ensure that no function appears in more than one split.

\subsection{Assembly and Source Normalization}

Assembly code is normalized to remove superficial lexical variation while preserving instruction semantics. Specifically, we apply the following preprocessing steps:
\begin{itemize}
    \item Removal of comments and instruction addresses.
    \item Stripping of register prefixes (e.g., \texttt{\%rax} $\rightarrow$ \texttt{rax}).
    \item Normalization of whitespace and punctuation.
    \item Conversion of hexadecimal constants to decimal form.
    \item Replacement of instruction addresses with symbolic placeholders (e.g., \texttt{[INST-1]}).
\end{itemize}

Source code is stripped of header inclusions and reformatted into a canonical style. These normalization steps reduce syntactic noise and ensure that retrieval emphasizes functional similarity rather than surface-level matching.

\subsection{Assembly Embedding and Indexing}

Each assembly function $A_i$ is encoded into a dense vector representation $h(A_i) \in \mathbb{R}^d$ using the encoder component of NOVA~\cite{jiang2023nova}, a pretrained foundation model designed for assembly code understanding. NOVA employs functionality contrastive learning and optimization contrastive learning to encourage embeddings of functionally equivalent code and to organize representations across optimization levels.

The final representation for each function is obtained by taking the mean of all instruction token embeddings. All embeddings are precomputed and indexed using FAISS, enabling efficient similarity search during retrieval.

\subsection{Similarity Computation and Category-Aware Re-ranking}

Given a target assembly function $A_t$ with embedding $h(A_t)$, similarity to each corpus function is computed using Cross-domain Similarity Local Scaling (CSLS), which mitigates the hubness problem in high-dimensional embedding spaces.

To further promote semantic relevance, we apply a category-aware re-ranking strategy. If the functional category of a candidate exemplar does not match that of the target function, its similarity score is downweighted by a penalty factor $\alpha$. This adjustment biases retrieval toward semantically related examples while allowing structurally similar cross-category exemplars to be selected.

The top-$k$ exemplars with the highest adjusted similarity scores are selected to form the retrieval context.

\subsection{Retrieval Procedure}

Retrieval is performed deterministically for all experiments. Given a target assembly function, its embedding is computed, similarity scores are calculated using CSLS, category-aware penalties are applied, and the top-$k$ assembly--source pairs are selected. The retrieved exemplars are ordered by decreasing similarity and inserted into the prompt as in-context demonstrations. Unless otherwise specified, we use $k = 5$.

\section{Optimization Flag Discovery}
\label{appendix:flags}

This appendix describes how compiler optimization flags used in the optimization-aware variant (ICL4D-O) are identified.

Modern compilers expose a large number of fine-grained optimization flags, many of which are interdependent and architecture-specific. To focus on optimizations that meaningfully affect emitted assembly in practice, we empirically identify which flags are active in our dataset.

Given a random subset of functions from the corpus, we extract a candidate list of optimization flags from GCC and Clang documentation. For each flag, we compile the same source code twice, once with the flag enabled and once with it disabled, while keeping all other compilation conditions fixed. The resulting assembly outputs are compared token-wise. A flag is considered active if enabling or disabling it changes the instruction sequence, register allocation, or control-flow structure.

To reduce the search space, we apply a binary search strategy by grouping related flags and iteratively testing subsets. This process yields a ranked list of optimization flags based on how frequently they affect assembly generation across sampled functions.

Based on this analysis, we select the most frequently active flags for use in optimization-aware prompting, including \texttt{-fomit-frame-pointer}, \texttt{-ftree-ter}, \texttt{-fipa-pure-count}, \texttt{-fcrossjumping}, and \texttt{-ftree-coalesce-vars}.

\section{Prompt Design for In-Context Decompilation}
\label{appendix:prompts}

This appendix describes the prompt design used in both variants of ICL4Decomp: 
retrieval-based in-context decompilation (ICL4D-R) and optimization-aware in-context decompilation (ICL4D-O).

\subsection{Prompt Design for Retrieved-Exemplar In-Context Decompilation (ICL4D-R)}
\label{appendix:prompt_icl4dr}

In ICL4D-R, the prompt is constructed by concatenating a small number of retrieved
assembly--source exemplars followed by the target assembly function and an explicit
decompilation instruction.

Each exemplar consists of an alternating pair of assembly code and its corresponding
high-level source implementation. Retrieved exemplars are ordered by decreasing
semantic similarity to the target assembly function. The prompt follows a consistent
structured format:
\begin{tcolorbox}[promptstyle, title={Example Prompt}]
\small
\texttt{[Example 1]}\\
\texttt{Assembly:} \textit{<A\_1>}\\
\texttt{Source:} \textit{<S\_1>}\\[2pt]
\texttt{[Example 2]} \ldots \textit{<A\_2, S\_2>}\\[2pt]
\texttt{[This is the assembly:]} \textit{<A\_t>}\\
\texttt{What is the source code?}
\end{tcolorbox}

This structured formatting exposes the language model to concrete instruction-to-structure
correspondences before generation, allowing it to implicitly adapt to the compiler style
and optimization level of the target function. Unless otherwise specified, we use
$k = 5$ retrieved exemplars. All other generation settings remain fixed across experiments.

\subsection{Prompt Design for Optimization-Aware In-Context Decompilation (ICL4D-O)}
\label{appendix:prompt_icl4do}

ICL4D-O augments the exemplar-based prompt with optimization-aware contextual guidance
that encodes compiler transformation semantics in natural language.

Each compiler optimization flag is associated with a rule descriptor consisting of four
components: (i) the optimization flag name, (ii) a natural language description of the
transformation, (iii) an illustrative source-level example, and (iv) a decompilation
hint explaining how to reason about the transformation during code reconstruction.

The prompt is extended with an \emph{Optimization Context} section that precedes the
target assembly code. This section informs the model that certain source-level constructs
may be absent or transformed due to semantics-preserving compiler optimizations, and
should be reconstructed accordingly during decompilation.

An example optimization-aware prompt follows:
\begin{tcolorbox}[promptstyle, float=ht, title={Example Prompt for \texttt{-ftree-coalesce-vars}}]
\small
\textbf{Optimize options instructions}\\
The binary you are decompiling may have been compiled with the GCC/Clang option 
\texttt{-ftree-coalesce-vars} (Tree SSA Variable Coalescing).\\[4pt]
-- This optimization merges multiple variables into a single register or memory location. 
It identifies variables that are copies of each other (e.g., \texttt{int b = a;}) or whose lifetimes do not overlap.\\
-- The goal is to eliminate redundant \texttt{mov} instructions and reduce register pressure, 
resulting in smaller and faster code.\\
-- As a result, intermediate variables from the source code may be completely absent in the final assembly; 
the logic is performed directly on the coalesced register.\\[4pt]
\textbf{Illustrative Source:}\\
\texttt{int coalesce\_example(int x, int y) \{}\\
\texttt{\ \ int a = x + 5;}\\
\texttt{\ \ int c = y - 2;}\\
\texttt{\ \ int b = a;}\\
\texttt{\ \ int d = b;}\\
\texttt{\ \ int e = c;}\\
\texttt{\ \ return d + e;}\\
\texttt{\}}\\[4pt]
\textbf{Decompilation Hint:}\\
Trace data flow from inputs to the return expression; 
do not create redundant temporary variables for copies (\texttt{b, d, e}) 
since they have been optimized away.\\
\texttt{[This is the assembly:]} \textit{<A\_t>}\\
\texttt{What is the source code?}
\end{tcolorbox}

When both retrieved exemplars and optimization rules are used, the optimization context
is inserted before the target assembly, while the exemplar demonstrations remain unchanged.
The remainder of the generation process is identical to that of ICL4D-R.


\subsection{Real-world Scenarios}

\begin{tcolorbox}[promptstyle, float=ht, title={LLM Judge Prompt}]
\small
\textbf{LLM-as-a-Judge evaluation}\\
Evaluate the quality of the recovered code against ground truth.\\
Score (0–100): 70\% semantic correctness, 20\% compilability, 10\% readability.\\[6pt]

\texttt{Ground Truth:}
\texttt{```c}
\textit{<GT\_CODE>}
\texttt{```}

\texttt{Candidate:}
\texttt{```c}
\textit{<PRED\_CODE>}
\texttt{```}\\[6pt]

\texttt{Return JSON only:}
\texttt{\{"score": <0-100>, "verdict": "<perfect|good|partial|poor|wrong>", "reason": "<short reason>"\}}
\end{tcolorbox}

\begin{table}[t]
\centering
\small
\caption{LLM-as-a-Judge scores grouped by dataset (HumanEval and ExeBench pending).}
\label{tab:llm_judge_grouped_by_dataset}
\resizebox{2\columnwidth}{!}{
\begin{tabular}{@{}lccccc ccccc ccccc@{}}
\toprule
\multirow{2}{*}{\textbf{Method}} 
& \multicolumn{5}{c}{\textbf{GitHub2025}} 
& \multicolumn{5}{c}{\textbf{HumanEval}} 
& \multicolumn{5}{c}{\textbf{ExeBench}} \\
\cmidrule(lr){2-6} \cmidrule(lr){7-11} \cmidrule(lr){12-16}
& Overall & O0 & O1 & O2 & O3
& Overall & O0 & O1 & O2 & O3
& Overall & O0 & O1 & O2 & O3 \\
\midrule
LLM4Decomp-1.3B
& 10.23 & 12.85 & 8.78 & 7.84 & 8.71
&  &  &  &  & 
&  &  &  &  &  \\

DeepSeek v3.2
& 26.65 & 32.44 & 23.33 & 21.05 & 23.51
&  &  &  &  & 
&  &  &  &  &  \\

Ours
& 35.63 & 40.16 & 35.98 & 34.62 & 30.48
&  &  &  &  & 
&  &  &  &  &  \\

\bottomrule
\end{tabular}
}
\end{table}

\end{document}